# Optical visualization of the enhanced spin Hall effect in bismuth doped silicon

Taiki Nishijima[1], Yakun Liu[2], Dushyant Kumar[2], Kyusup Lee[2], Fabien Rortais[1], Syuta Honda[3], Yuichiro Ando[1,†], Ei Shigematsu[1], Ryo Ohshima[1], Hyunsoo Yang[2] and Masashi Shiraishi[1]

1. Department of Electronic Science and Engineering, Kyoto University, Kyoto, Kyoto 615-8510, Japan
2. Department of Electrical and Computer Engineering, National University of Singapore, Singapore 117576, Singapore
3. Department of Pure and Applied Physics, Kansai University, Suita, Osaka 564-8680, Japan

Abstract

Direct visualizations of spin accumulation due to the enhanced spin Hall effect (SHE) in bismuth (Bi) - doped silicon (Si) at room temperature are realized by using helicity-dependent photovoltage (HDP) measurements. Under application of a dc current to the Bi-doped Si, clear helicity-dependent photovoltages are detected at the edges of the Si channel, indicating a perpendicular spin accumulation due to the SHE. In contrast, the HDP signals are negligibly small for phosphorus-doped Si. Compared to a platinum channel, which has a large spin Hall angle, more than two-orders of magnitude larger HDP signals are obtained in the Bi-doped Si.

Silicon (Si) is an attracting material for spintronics because it enables a good spin coherence due to its weak spin-orbit coupling (SOC) [1–8]. A steady progress has been made to realize the practical Si-based spin devices, and recently, the spin MOSFET and the spin logic gates have been demonstrated at room temperature [9–11]. In these devices, however, ferromagnetic contacts with high interfacial resistance were employed to solve the spin resistance mismatch, resulting in the increment of energy consumption [12]. Degradation of thermal stability of the device is also expected, because of easy formation of the silicides. Although the creation and detection of spin polarization via the spin Hall effect (SHE) and the inverse spin Hall effect (ISHE) without ferromagnetic materials are alternative ways, they have been regarded inherently difficult in Si because of the weak SOC. Accordingly, the usage of Si has not yet been fully explored as active spin devices where the spin function, typically supported by the spin Hall effect (SHE), is an essential. Therefore, if the enhancement of the SOC in Si is realized, we can add a new functionality to the Si-based spin devises and all-Si spin devices consisting of a Si injector, a Si detector, and a Si channel can be realized.

Recently, we reported modulation of SOC strength in Si by doping heavy atoms [13]. Bismuth (Bi) is a heavy element and also acts as a donor in the Si [14], which can provide both carriers and ability of spin function via the SOC. A clear change from the weak localization (WL) to the weak anti localization (WAL) was demonstrated by Bi doping into the Si fabricated by the ion implantation technique, indicating an enhancement of the SOC. The proposed way has a good compatibility with the present manufactures and enables coexistence of strong SOC with a long spin lifetime in the same device. However, the WL and WAL measurements was limited only at low temperature and the investigation of the enhanced SOC-driven SHE effect is desirable.

In this study, we demonstrate direct visualizations of spin accumulation due to the enhanced SHE in Bi-doped Si at room temperature by using helicity-dependent photovoltage (HDP) measurements [15,16]. Under application of a dc current to a Bi-doped Si channel, clear helicity-dependent photovoltages are detected at the edges of the Si channel, indicating the spin accumulation with out-of-plane spin textures as shown in Fig. 1(a). In contrast, the HDP signals are negligibly small for a phosphorus (P)-doped Si channel. The magnitude of HDP signals in the Bi-doped Si is more than two-orders of magnitude larger than that in a platinum (Pt) channel, a representative material with a large spin Hall angle.

In the experiments, we used the same Bi-doped Si channel as in the previous study [13]. Samples were fabricated using an intrinsic silicon-on-insulator substrate with a 100-nm-thick Si channel. The Si channel was doped with P by ion implantation to facilitate the following HDP measurements by ensuring the electrical conductivity. The rapid thermal annealing was carried out at 900 °C for 1 second. The substrate only with P dopants was used as a reference sample, which is denoted as "P-Si sample". After activation of the P dopants, Bi was additionally doped and annealed at 600 °C for 30 mins [14]. Concentration of P and Bi calculated by using the Transport of Ions in Matter program were $1\times10^{20}$ and

$5\times10^{19}$ cm$^{-3}$, respectively, indicating a formation of degenerate Si channel. A depth profile of each dopant was almost uniform except for the top 20 nm region, by using the multi-step ion implantation with different acceleration energies. The detailed procedures are described elsewhere [13]. The P and Bi co-doped sample is denoted as "Bi-Si sample". The resistivity and carrier concentration of the Si channel for the P-Si sample were $2.4\times10^{-3}$ Ωcm and $4.0\times10^{19}$ cm$^{-3}$, respectively, at 300 K. For Bi-Si sample, those quantities were found to be $4.2\times10^{-3}$ Ωcm and $5.0\times10^{19}$ cm$^{-3}$, respectively, indicating a part of the implanted P and Bi ions was activated [13]. Hall measurements revealed that the electron mobility was decreased from 64 to 18 cm$^2$V$^{-1}$s$^{-1}$ at 300 K by Bi implantation, which suggests the enhancement of electron scattering probability by Bi atoms [13]. The optical microscope image of the fabricated Hall bar is shown in Fig.1(b). The electron beam lithography and argon ion (Ar$^+$) milling were used to pattern the Si channel into the Hall bar structure. We etched the top 50 nm of the Si layer to use the uniformly doped region as a channel. The experimental setup is shown in Figs. 1(b) and 1(c). In the HDP measurements, the light circular polarization was modulated by a photo elastic modulator (PEM) at a fixed frequency of 50 kHz. A laser light (wavelength $\lambda$ = 662 nm, power: $P_{op}$ = 6 mW) was focused at normal incidence onto the Si channel using an objective lens (×80). The laser spot size was ~1.5 μm in diameter which determines the spatial resolution. The photo current is expected to be spin polarized by irradiation of the circularly polarized light. When the current driven spin accumulation by the SHE is taken place in the channel, the number of occupied and empty states in the conduction band also has spin polarized. In that situation, the amount of photo current by circularly polarized light depends on the amount of spin accumulation. We measured the helicity dependent local resistivity induced by the light-spin coupling by using lock-in technique. The detailed procedures of the HDP measurements are described elsewhere [15].

Figures 2(a)-2(e) show the scanning HDP data in the Si channel of the Bi-Si sample. The black bold arrows indicate the direction of the applied charge current. A constant background voltage detected in the whole scanning area was subtracted from the raw data (see Fig. S1 in the supplementary data [17]). The line scans of the HDP signals obtained by averaging the HDP voltage along the $y$-direction are also displayed in the bottom panels. No clear signals were recognized at $J_{DC}$ = 0 A/m$^2$ (Fig. 2(c)). Upon application of the charge current, clear HDP signals were obtained at the edges of the channels and the signs of the signals were opposite at two edges. Reversing the bias current direction caused the HDP signal to switch its sign, typical features of the spin accumulation generated by the SHE, the quantization axis of which is along ±z direction in Fig. 1(a). In contrast, no clear HDP signals were obtained at the edges of the Si channel for the P-Si sample as shown in Figure 3, even though comparable $J_{DC}$ was applied to the Si channel. The clear difference in HDP signals between the Bi-Si and P-Si samples indicates an enhancement of the SHE by Bi doping even though atomic composition of Bi atoms is only 0.1 %, smaller than previous study which used copper-Bi alloys [18].

Hereafter, we consider the origin of the enhancement of the HDP signal by Bi doping. The HDP

voltage can be expressed as following equation:

$$V_{\mathrm{HDP}} = \left[\left(\frac{m_{\mathrm{e}}}{\mu_{\mathrm{e}}\left(\frac{L}{wd}N_{\mathrm{e}}+\Delta N_{\mathrm{op}}\right)}\right)^{-1} + \left(\frac{m_{\mathrm{h}}}{\mu_{\mathrm{h}}\left(\frac{L}{wd}N_{\mathrm{h}}+\Delta N_{\mathrm{op}}\right)}\right)^{-1}\right]\frac{wdJ_{\mathrm{DC}}}{e}, \quad (1)$$

where, the suffix "e" and "h" denote the electron and hole, respectively. $m$, $\mu$, and $e$ are the effective mass, the mobility of carriers and the elementary charge, respectively. $N$ and $\Delta N_{\mathrm{op}}$ are the initial and net of optically generated carrier densities. $\Delta N_{\mathrm{op}}$ is expressed as following equation:

$$\Delta N_{\mathrm{op}} = \tau_{\mathrm{e-h}}\frac{P_{\mathrm{op}}}{h\nu}(1-R)(1-\exp[-(\alpha_{\mathrm{L}}-\alpha_{\mathrm{R}})d]), \quad (2)$$

where, $\alpha_{\mathrm{L}}(\alpha_{\mathrm{R}})$ is the absorption coefficient of the left (right)-handed circularly polarized light, $h$ is the plank constant, $\nu$ is the frequency of the light, $R$ is the reflectance of the sample, $d$ is the thickness of the channel and $\tau_{\mathrm{e-h}}$ is the lifetime of the photo excited carriers, respectively. Generally, $\alpha_{\mathrm{L}}$ and $\alpha_{\mathrm{R}}$ are identical in Si because of negligible circular dichroism. However, $\alpha_{\mathrm{L}} - \alpha_{\mathrm{R}}$ is expected to become nonzero when the following two conditions are both satisfied: (i) an local imbalance of the spin polarization, i.e., the spin accumulation at the boundary of the Si channel, (ii) a finite net spin polarization induced by the helicity-dependent selection rule for the interband optical transition. The spin Hall angle of P-doped Si sample is expected to be negligibly small because P and Si are both light elements and the carrier type is electron [19]. Therefore, the spin accumulation in the Bi-Si sample is mainly attributed to the SHE related to Bi atoms. There are two possible mechanisms of the enhanced SHE by Bi doping. One is an enhancement of the extrinsic SHE by introducing the additional electron scattering centers. The electron mobility was actually decreased from 64 to 18 cm$^2$V$^{-1}$s$^{-1}$ by Bi doping [13]. The other possible origin is modulation of Si band structure which introduces the intrinsic SHE. Further investigation is needed to reveal the dominant factor for the condition (i).

In contrast, the required condition (ii) is related to the band structure of Si. Because the photon energy of the light (1.87 eV) is larger than the indirect band gap (1.17 eV) of Si, but is smaller than the direct band gap at Γ point (3.4 eV), the carriers were generated via the indirect transition assisted by phonon emission or absorption [20]. Although the selection rule of indirect optical transition in Si was theoretically and experimentally investigated, the spin polarization of it is estimated to be negligibly small when the photon energy is sufficiently large compared to the band gap [20–24]. Therefore, if the band structure of the Bi-Si sample is similar to that of the pure Si, a negligible $\alpha_{\mathrm{L}} - \alpha_{\mathrm{R}}$ is expected even though a sizeable spin accumulation in Si is generated via the SHE.

To investigate the effect of Bi doping on the band structure of Si, the first principle calculations were carried out. We calculated the band structure of Bi- or P-doped Si for various impurity concentration and location. An impurity Bi or P atom was added at an interstitial or substitutional position. Number of lattice points is 54. The detailed calculation procedures are described in the supplementary data [17]. The unit cells of crystal structures of Bi and/or P-doped Si for the calculation are shown in Fig. 4(a). Two typical cases were considered; The impurity atoms are doped at substituted or interstitial positions. The band structures without and with the SOC were compared to reveal the effect of the SOC. First, we substituted one of Si atoms with a Bi atom as shown in Fig. 4(a), corresponding to a doping concentration of $9.1 \times 10^{20}$ cm$^{-3}$. The band structure is shown in Fig. 4(c). The energy gap, $\Delta E_{SO}$, between the light hole (LH) or heavy hole (HH) bands and the split off (SO) band was calculated to be 0.0724 eV, slightly larger than that of the P-doped Si (0.0465 eV, Fig. 4(d)) and the pure Si (0.0472 eV, see the supplementary data), indicating enhancement of the SOC. However, enhancement of $\Delta E_{SO}$ is not sufficient to suppress the optical transition from the SO band to the CB because the photon energy of the laser light was considerably larger than the energy gap between the SO band to the CB. In such a case, the net spin polarization of electrons excited from HH, LH and SO bands is calculated to be almost 0 as schematically shown in Fig. 4(b). Interstitial Bi and P atoms also showed small $\Delta E_{SO}$ as shown in Figs. 4(f) and 4(g).

Next, we focus on the CB. The band structure was complicated compared to that of the pure Si (see the supplementary data), because of hybridization of orbitals between Si and impurity atoms. It is noted that a clear difference in the CB between without and with SOC cases was confirmed only for the interstitial Bi atom shown in Fig. 4(f), indicating non-negligible contribution of p- or d-orbitals of the Bi atoms. The density of states (DOS) of Bi atoms as a function of energy for substituted and interstitial Bi atoms in Si is shown in Fig. 4(e) and 4(h), respectively. For the interstitial Bi atoms, large DOS of the p-orbital were located around the bottom of the CB. In that case, the hybridization of p-orbitals of the Bi atoms and s-orbitals of the Si atoms is expected, resulting in introduction of the orbital angular momentum into the CB as well as introduction of SOC. As a result, the spin polarization due to the selection rule might be enhanced both for direct and indirect optical transitions as schematically shown in Fig. 4(b). Furthermore, an enhancement of the intrinsic SHE is also expected. Although optical transition from p-orbital to p-orbital is forbidden because of parity of wave function, hybridization with s-orbital and contribution of phonon might allow such an optical transition. The modulation of the CB was confirmed even for the Si with Bi concentration of $4.9 \times 10^{19}$ cm$^{-3}$, close to the experimental one. Because the band-structure modulation by donors or acceptors, well known as the "band gap narrowing", was confirmed even for the relatively low impurity concentration down to $10^{17}$ cm$^{-3}$ [25,26], the calculation results in Fig. 4 are reasonable and qualitatively explain the enhancement of HDP signals by Bi doping.

Finally, we discuss the magnitude of $|\Delta V_L|$ signal shown in Fig. 2. Because $|\Delta V_L|$ depends on $P_{op}$, and $J_{DC}$, here we focus on $|\Delta V_L|/(J_{DC}P_{op})$. The $|\Delta V_L|/(J_{DC}P_{op})$ value of the Bi-Si sample was $1.5 \times 10^{-11}$ m$^2$/A$^2$ for $J_{DC} = 4.2 \times 10^9$ A/m$^2$, which is more than two-orders of magnitude larger than that of the Pt channel reported so far [15]. Such a large $|\Delta V_L|/(J_{DC}P_{op})$ is mainly attributed to small $N_e$, and long $\tau_{e-h}$ owing to the indirect interband optical transition.

In summary, we demonstrated direct visualizations of spin accumulation due to the enhanced SHE

in Bi-doped Si at room temperature. The HDP measurements showed a clear difference in the signal magnitude between the P-Si and Bi-Si samples, indicating an enhancement of the SHE by Bi doping. First principle calculation revealed that interstitial Bi atoms modulate the conduction bands and induce a finite orbital angular momentum, resulting in increment of spin polarization in the selection rule for the interband optical transition. Compared to the Pt channel, more than two-orders of magnitude larger HDP signals were obtained. The Bi doping into Si provides two abilities to the Si-based spin devices: an optical injection and detection of spin accumulation in Si and an efficient spin-charge interconversion via the SHE and the ISHE.

ACKNOWLEDGMENTS

This work was supported by JSPS (KAKENHI No. 16H06330, No. 18H01465 and No. 19H02197) and SpOT-LITE program (A*STAR Grant, A18A6b0057) through RIE2020 funds.

FIGURE CAPTIONS

FIG. 1. (a) A schematic of spin accumulation induced by the spin Hall effect with positive spin Hall angle for helicity dependent photovoltage (HDP) measurements. (b) An optical image of the Hall bar device with a representation of the electrical connections. The HDP photovoltage was acquired by a lock-in amplifier. (c) An experimental setup of the HDP measurements. The laser light with a wavelength of 662 nm was focused at normal incidence onto the sample. The helicity was modulated using a photo elastic modulator (PEM) with a fixed frequency of 50 kHz.

FIG. 2. Spatial two-dimensional HDP maps in Bi-Si sample under bias currents of (a) $-4.2 \times 10^9$, (b) $-2.1 \times 10^9$, (c) 0, (d) $2.1 \times 10^9$, (d) $4.2 \times 10^9$ A/m$^2$. Black dashed lines indicate the edges of the device confirmed by optical reflection measurements. Black bold arrows show the applied bias current direction. Line scans of HDP, averaged along the y-direction, are displayed in the bottom panels.

FIG. 3. Spatial two-dimensional HDP maps in the P-Si sample under bias currents of (a) $-4.2 \times 10^9$, (b) $-2.1 \times 10^9$, (c) 0, (d) $2.1 \times 10^9$, (d) $4.2 \times 10^9$ A/m$^2$. Line scans of HDP, averaged along the y-direction, are displayed in the bottom panels.

FIG. 4. (a) Schematics of the Si crystal structure for the first principle calculations. (b) A sketch of optical transitions allowed with right-handed circular polarized light $\sigma^-$ at $\Gamma$ point of Si. Each state is written in blue. The red arrows stand for allowed transitions, whose probabilities are proportional to the product of the numbers next to the arrows. Hybridization of p- or d-orbitals of Bi atoms with s-orbital of Si might increase the splitting and create additional states with finite orbital angular momentum at the conduction band which increase the spin polarization of optical transition as shown in green lines. Although indirect optical transition process is more complicated because of contribution of phonon, similar enhancement of spin polarization can be expected. (c, d) Band structures of Si with substituted (c) Bi or (d) P atoms. We calculated the band structures without and with SOC, which are shown in red and black lines, respectively. (e) The density of states (DOS) of Bi atoms as a function of energy for substituted Bi atoms in Si. (f, g) Band structures of Si with interstitial (f) Bi or (g) P atoms. (h) DOS of interstitial Bi atoms in Si as a function of energy.

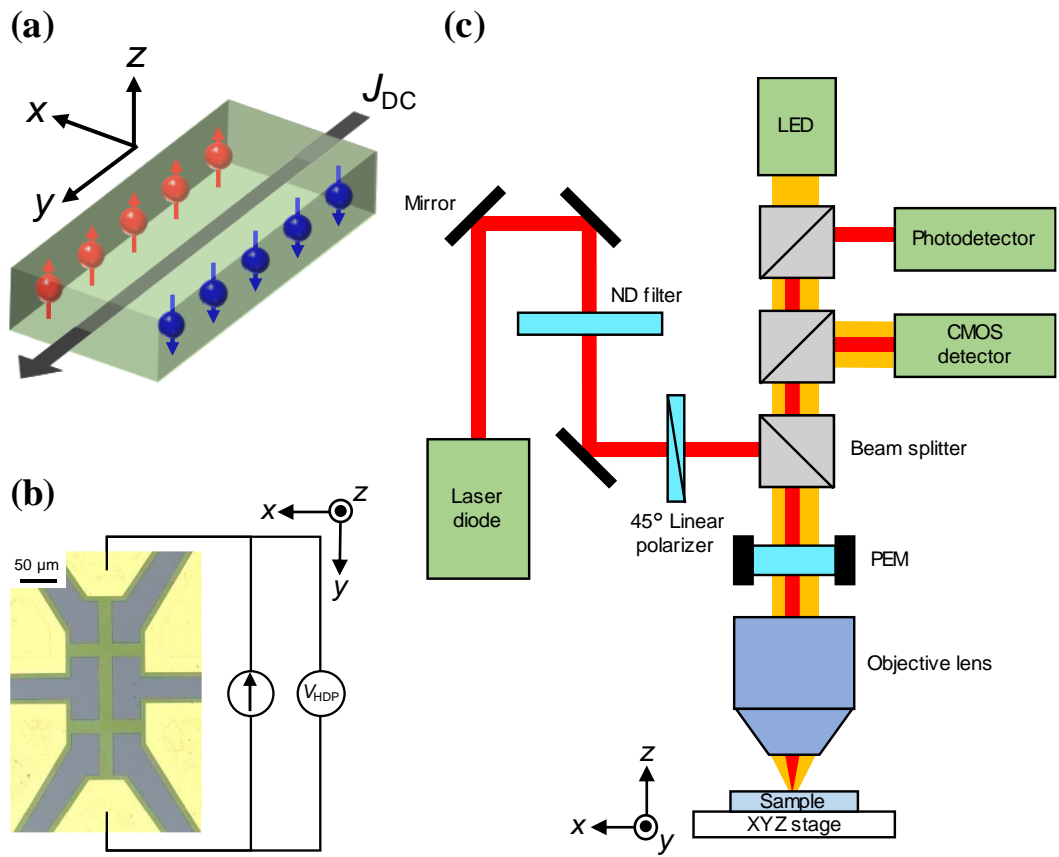

Fig.1 Nishijima et al.,

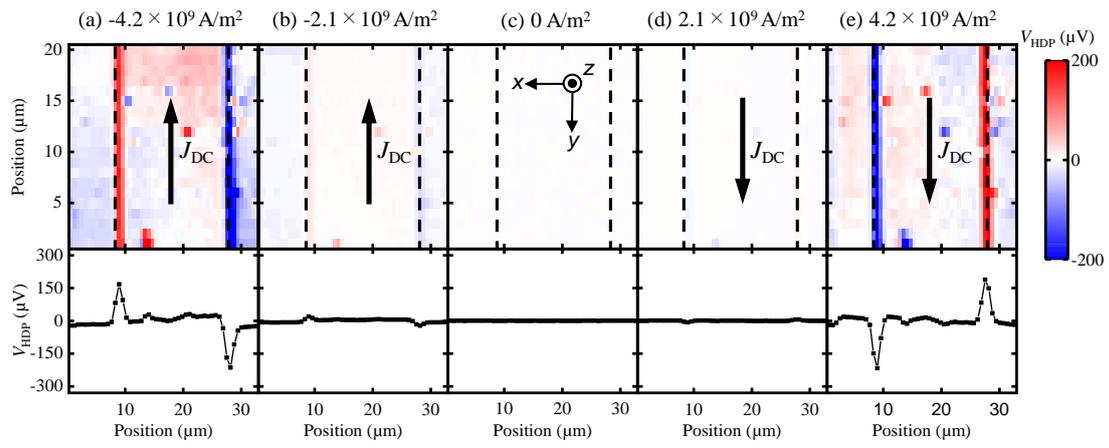

Fig.2 Nishijima et al.,

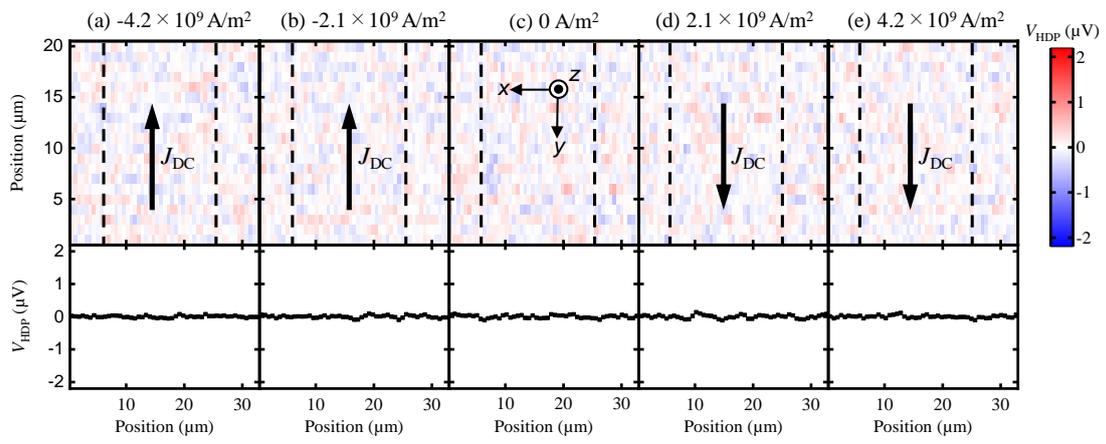

Fig.3 Nishijima et al.,

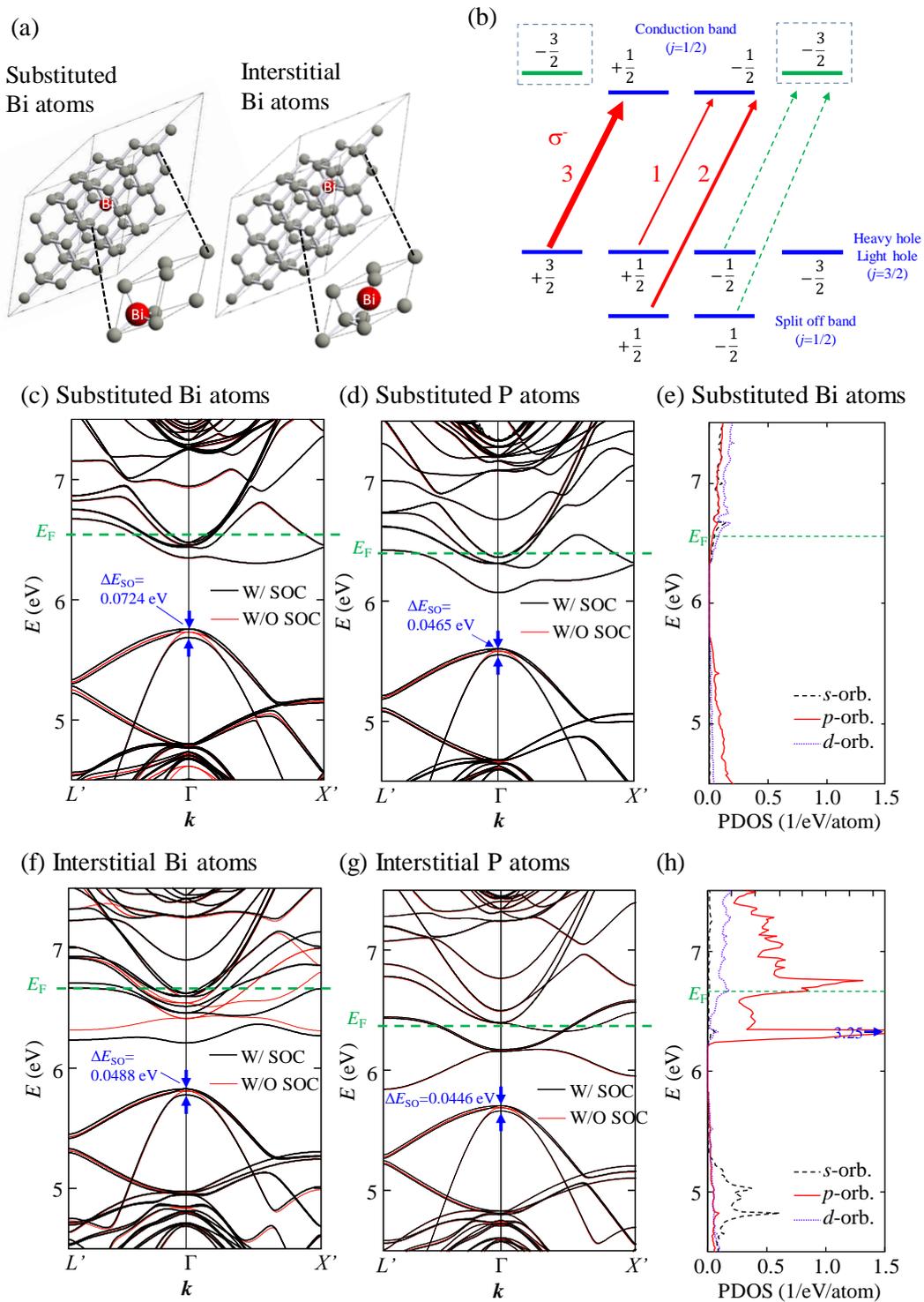

Fig.4 Nishijima et al.,